\documentclass[%
 reprint,
 amsmath,amssymb,
prb,
]{revtex4-1}

\usepackage{textgreek} 
\usepackage{graphicx}
\usepackage{dcolumn}
\usepackage{bm}
\usepackage[tight,footnotesize]{subfigure}
\usepackage{hyperref}
\usepackage[mathlines]{lineno}
\pdfoutput=1 

\begin{document}

\preprint{APS/123-QED}

\title{Saturated Low-Temperature Conductivity \\
in Ultrafast Semiconductor Nanocomposites}

\author{W. Zhang}%
\email{weidong.zhang@wright.edu}
\author{E. R. Brown}
\email{elliott.brown@wright.edu}

\author{M. Martin}%
\affiliation{%
 Terahertz Sensor Laboratory, Depts. of Physics and Electrical Engineering,\\ Wright State University, Dayton, OH, USA, 45435
}%



\date{June 10, 2013}

\begin{abstract}
This article presents studies on low-field electrical conduction in the range 4-to-300 K for a ultrafast material: InGaAs:ErAs grown by molecular beam epitaxy. The unique properties include nano-scale ErAs crystallines in host semiconductor, a deep Fermi level, and picosecond ultrafast photocarrier recombination.  As the temperature drops, the conduction mechanisms are in the sequence of thermal activation, nearest-neighbor hopping, variable-range hopping, and Anderson localization. In the low-temperature limit, finite-conductivity metallic behavior, not insulating, was observed. This unusual conduction behavior is explained with the Abrahams scaling theory. 
\end{abstract}

\maketitle
Terahertz devices can be fabricated from a new class of semiconductors such as InGaAs:ErAs.\cite{r1, r2, r3, r4} Given the bandgaps of host semiconductors InGaAs $\sim$ 0.8 eV, ultrafast photoconductors made from this material are compatible with the 1.55-micron EDFA fiber laser technology. InGaAs: ErAs was grown on semi-insulating InP substrates with molecular beam epitaxy. The impurities - both rare earth erbium (Er) and beryllium (Be) atoms- were deposited simultaneously along with the In0.53Ga0.47As layers. Previous studies revealed that erbium bonds covalently with arsenic forming nano-scale crystallite ErAs islands. \cite{r1, r2} These ErAs islands have atomically abrupt interfaces with their host semiconductor. 
As a result, there were few defects in this nanocomposite. The photocarrier recombination time was measured as short as 1 ps. \cite{r3}
The sizes of the ErAs particles were in the range from 30nm x 4nm to as small as 2nm x 2nm according to The Tunneling Electron Microscopy (TEM) images. TEM images also showed the ErAs nanoparticles were randomly positioned in the host semiconductor.
Although there have been plenty of device implementations,\cite{r1, r2, r3, r4} the electrical conduction mechanisms of such kind of ultrafast composite materials, especially at low temperatures, remains to be understood. In this communication, we report the cryogenic characterization of InGaAs:ErAs as well as the modeling of observed results. 


For the InGaAs:ErAs sample characterized in this research, the epitaxial InGaAs layer was 2 $\mu$m thick. It follows by a 0.1 $\mu$m InAlAs buffer layer and then became latticed-matched to a semi insulating InP substrate. The fraction of erbium was about 0.3\%, and the Be concentration was $5\times 10^{18}$ cm\textsuperscript{-3}. The Beryllium ions as acceptors provide compensation. According to Hall measurements, the sample was n-type; the resistivity was 13.1 $\Omega$-cm; and the room-temperature mobility was 384 cm\textsuperscript{2}/V-s.

A second sample was consisted of a 1-micron thick epitaxial layer of MBE-grown undoped epitaxial InGaAs lattice-matched to a semi-insulating InP substrate.  Because of its lack of intentional doping, this acted as the control sample in our experiments.

\begin{figure}
	\centering
	\includegraphics[scale=0.45]{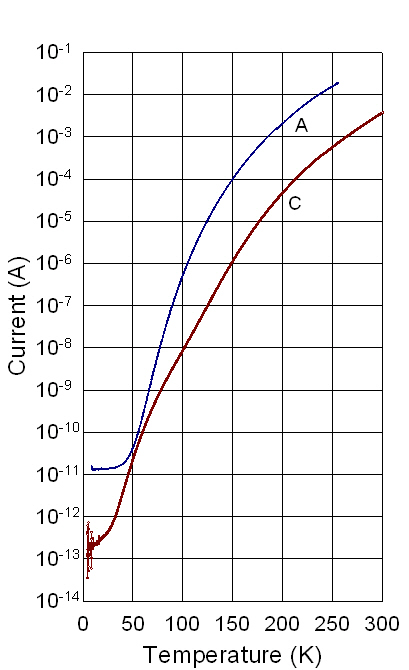} 
	\caption{Current vs  temperature for A: InGaAs:ErAs; and C: an undoped InGaAs.}
	\label{fig:ABCcurves}
\end{figure}

      The samples were mounted using low-temperature thermal grease (Apiezon N) on the copper cold finger of a Gifford-McMahon close-cycle-He refrigerator. The grease accommodated the different thermal expansions between the copper and semiconductor substrate, and also provided a significant level of thermal conductivity at the lowest temperatures.  The temperature on the cold finger was measured with a calibrated DT-670B-CU silicon diode, and then recorded with a Lakeshore temperature controller. The thin wires to the samples were thermally and electronically insulated, and brought out of vacuum with BNC vacuum feed-through connectors. The measurements were performed during both the ramp-down in temperature (rate-limited by the refrigerator) and the ramp-up in temperature (after turning off the compressor).   The sample electrical characteristics were almost identical in both directions, confirming good thermal contact without hysteresis. 
      
      The InGaAs:ErAs sample was biased at 2.0 V with dry alkaline batteries for their low-noise characteristics. The DC current was measured with a Keithley 6487 picoammeter having the range 20 fA-20 mA.  The distance between electrodes was 9-$\mu$m for the InGaAs: ErAs.  Hence the maximum electric field was $\sim 2\times 10^{3}$ V/cm, which is considered to be low-field conditions.

\begin{figure}
	\centering
	\includegraphics[scale=0.38]{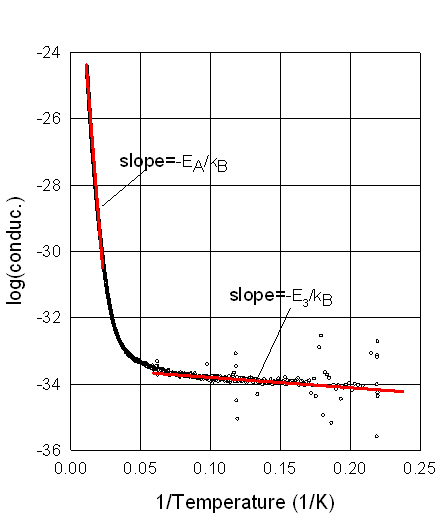} 
	\caption{The electrical conductance of the InGaAs control sample.}
	\label{fig:controlsample}
\end{figure}

    Current vs. temperature curve, labeled as A, is plotted in Fig. \ref{fig:ABCcurves}.  The curve drops quickly between room temperature and $\sim$100 K, but approaches an asymptotic level in the low-temperature limit. The asymptotic level, $\sim$13 pA for the InGaAs: ErAs is well above the minimum instrumental values: 20 fA for the Keithley 6487.  The possibility of temperature saturation by poor thermal conductance between the sample and the cold finger was also investigated, but ruled out, based on the behavior of the control sample, labeled as C in Fig. 1. The control sample was biased at 100 V with a CSI12001X power supply, and the current measured with a Keithley 6514 electrometer having the range 0.1 fA-20 mA. The distance between electrodes was $\sim$500 $\mu$m, hence the maximum electric field was approximately the same as the previous sample. The current in this sample dropped monotonically with decreasing temperature, and approached the level of 0.1pA at $\sim$5K.  The Arrhenius plot of the control sample (Fig. \ref{fig:controlsample}) is well fit in the low temperature range $\sim$56-90K by the activation energy $\sim$55.4 meV. We attribute this to a background impurity (probably carbon) in the undoped InGaAs control sample.  In the lowest temperature range, there is another Arrhenius fitting from which the activation energy is estimated $\sim$0.3meV (Fig. \ref{fig:controlsample}). Such small activation energy can be fit into 'Miller and Abrahams' nearest- neighbor hopping and has been seen in the impurity conduction of semiconductors. \cite{r7} In contrast, no thermally activated behavior was observed in the InGaAs: ErAs at the lowest temperatures.  Instead its saturated-current behavior shown in Fig. 1 is indicative of metallic transport.
\begin{figure}[h]
	\centering
	\includegraphics[scale=0.35]{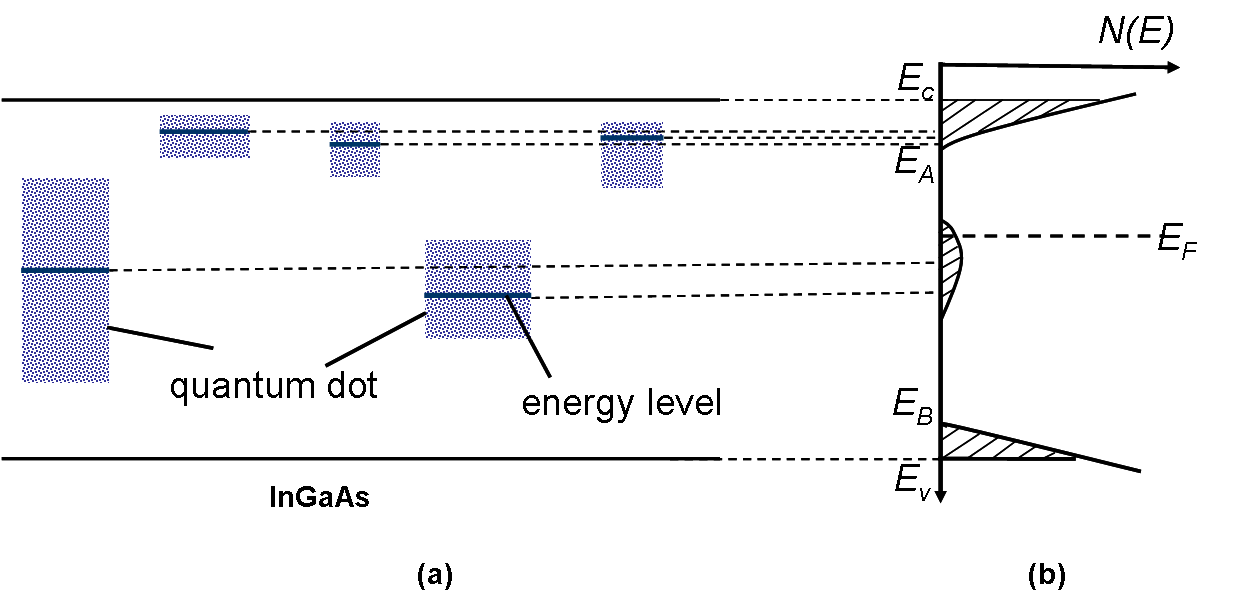} 
	\caption{(a) The nano-scale quantum dots and their energy levels; and (b) the Mott mobility edges. The cross hatching represents localized energy states.}
	\label{fig:quantumdot}
\end{figure} 

       To understand the electrical transport in these samples, a potential-well model was developed based on the nano-scale ErAs islands acting as quantum dots.\cite{r4}  A potential well is formed from the band discontinuities between the ErAs island and the InGaAs.   The energy levels of an isolated well become deeper or shallower as the size of quantum dot increases or decreases, respectively.  There are tens of thousands randomly distributed potential wells, and  their sizes are random too. By general quantum mechanical principles, the energy levels of this statistical ensemble expand into a quasi-continuous density-of-states (DoS), N(E). The Mott model with the host semiconductor setting the energy scale is an appropriate starting point (Fig. \ref{fig:quantumdot}). $E_{C}$ and $E_{V}$ are the conduction and valence band mobility-edge energies, respectively. \cite{r7} The states just below $E_{C}$ or just above $E_{V}$ are localized.  So are the states near the Fermi level $E_{F}$ . The energy states above $E_{c}$ are extended states where the conduction is by free-electrons via the semiconductor bands. With the help of Fig. \ref{fig:quantumdot}, the universal conduction mechanisms of Fig. \ref{fig:ABCcurves}, indexed (i)-(iv), are analyzed.\cite{r7}
       
(i) Near room temperature the conduction is created by thermal activation of the localized electrons into the extended states above $E_{c}$. The Arrhenius equation is,
\begin{equation}
G\propto \exp{\left( -\frac{E_{C}-E_{F}}{k_{B}T} \right)}
\label{eq:c1}
\end{equation}  

(ii) At somewhat lower temperature, the conduction is dominated by thermally-activated hopping of electrons in the localized states lying between mobility edges $E_{C}$ and $E_{A}$,                                                   
\begin{equation}
G\propto \exp{\left( -\frac{E_{A}-E_{F}+w_{1}}{k_{B}T} \right)}
\label{eq:c2}
\end{equation}                                                                                                                                            where $w_{1}$ is the island-to-island hopping energy. 
 
(iii) At yet lower temperatures, conduction is dominated by the thermally activated hopping among the localized states near the Fermi level $E_{F}$, 
\begin{equation}                                                                 G\propto \exp{\left( -\frac{w_{2}}{k_{B}T} \right)}                                                                 
\label{eq:c3}
\end{equation}
where $w_{2}$ is another island-to-island hopping energy.

(iv) At the lowest temperatures, conduction is mainly from the Mott variable-range hopping among the localized states near the Fermi level, 
\begin{equation}                                                                 G\propto \exp{\left( -\frac{B_{M}}{T^{\frac{1}{4}}} \right)}  
\label{eq:c4}
\end{equation}
where $B_{M}$ is a constant.

\begin{figure}
	\centering
	\includegraphics[scale=0.3]{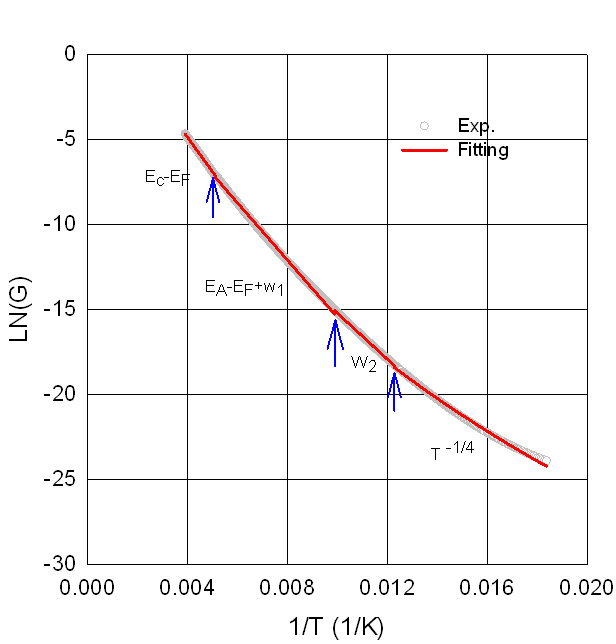} 
	\caption{Conductance vs temperature with the Mott model, InGaAs: ErAs sample. }
	\label{fig:sample1_14}
\end{figure}

    The conduction modes (i)-(iv) together with their fitting parameters are plotted in Fig. \ref{fig:sample1_14}. The parameters, $E_{C}-E_{F}$, $E_{A}-E_{F}+w_{1}$, $w_{2}$, and $B_{M}$ are listed in Table \ref{table:ta1}.  The Fermi level is located deeply inside the bandgap of InGaAs, and the conduction mechanisms (i)-(iv) are all from localized states.

\begin{table}[ht]
\caption{The fitting parameters of Mott model} 
\centering 
\begin{tabular}{c c c c } 
\hline\hline 
    $E_{c}-E_{F} (eV)$  & $E_{A}-E_{F}+w_{1} (eV) $ & $w_{2} (eV)$ & $B_{M}$ \\ [0.5ex] 
\hline 
 0.175 & 0.146 & 0.117 & 165.1\\ 
\hline 
\end{tabular}
\label{table:ta1} 
\end{table}
    
       Interestingly, the Mott model of Fig. \ref{fig:quantumdot} does not explain the observed asymptotic conductance at the low temperatures below 45 K.  For this, we need to consider the quantum scaling theory of the metal-insulator transition.  
       
\begin{figure}
	\centering
	\includegraphics[scale=0.43]{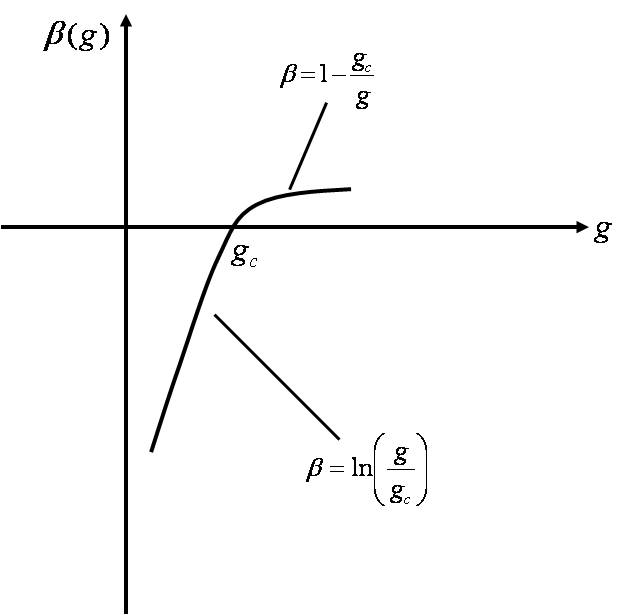} 
	\caption{The plot of $\beta(g)$ function and its behavior. \cite{r8} }
		\label{fig:scaling}
\end{figure}

(v) As the temperature drops to the lowest levels, the conductivity is in the metallic state of the transition. The Abrahams scaling equation is 
\begin{equation}
d\ln{(g)}/d\ln{(L)}=\beta(g)
\label{eq:Abrahams}
\end{equation}
where $g$ is the dimensionless conductance of a hypercube with the side length $L$. 
The conductivity is connected to $g$ through $\sigma=e^{2}g/\hbar L$, where $e$ is the electron charge, and $\hbar$ is the Planck constant.  
$g_{c}$ is the critical conductance at which $\beta(g)$ crosses zero (Fig. \ref{fig:scaling}). 
For the metallic side of the transition, $g>g_{c}$, and $\beta(g)$ is taken as, \cite{r8}
\begin{equation}
\beta(g)=1-\frac{g_{c}}{g}
\label{eq:Abrahams_metal_solu}
\end{equation}
The integration of the equations (\ref{eq:Abrahams}) and (\ref{eq:Abrahams_metal_solu}) yields  $g=g_{c}+L/l_{0}$ along with $l_{0}$ a constant. The following conductivity is then obtained, \cite{r9, r10}  
\begin{equation}
\sigma=\frac{e^{2}g_{c}}{\hbar L}+\frac{e^{2}}{\hbar l_{0}}
\label{eq:Abrahams_metal_solu_cond}
\end{equation}

Altshuler and Aronov argued that the length $L$ should be replaced with the interaction length $L_{int}$ near the metal-insulator transition,\cite{r10} $L_{int}=\sqrt{D\hbar/k_{B}T}$ , where $D$ is the diffusion constant and $k_{B}$ is the Boltzmann constant.  However, $D$ can also bring in temperature dependence through the Einstein relation.  After this correction, the conductance has a 1/3 power-law dependence on temperature along with a temperature-independent term that makes the conductance non-vanishing as T approaches zero: \cite{r9} 
\begin{equation}                                                                        \sigma=\frac{e^{2}}{\hbar}\left[ N(E_{F})g_{c}^{2}k_{B}\right]^{1/3}T^{1/3}+\frac{e^{2}}{\hbar l_{0}} 
\label{eq:Abrahams_metal_solu_cond_AA}
\end{equation}                                                                             where $N(E_{F})$ is the density of states at the Fermi level.

Away from the variable-range hopping conduction on the localized side of the transition, the scaling function $g<g_{c}$, and $\beta(g)$ is taken as \cite{r8},
\begin{equation}
\beta(g)=\ln{\frac{g}{g_{c}}}	
\label{eq:Abrahams_insulator_solu}
\end{equation}                                                                     
The integration of Eq.(5) and Eq.(9) yields $g=\exp{(-L/\xi)}/g_{c}$, where $\xi$ is the localization length. The conductivity is then given by:
\begin{equation}
\sigma=\frac{e^{2}}{\hbar g_{c} L_{in}}\exp{\left(-\frac{L_{in}}{\xi}\right)}
\label{eq:Abrahams_insulator_solu_conduc}
\end{equation}
Here $L$ is replaced with the inelastic diffusion length $L_{in}$, which is a power function of temperature, $L_{in}\propto T^{-b}$ or $L_{in}=B T^{-b}$ with $B$ being the proportionality coefficient. \cite{r11}  Then the conductance for this mechanism is given by
\begin{equation}
\sigma=\frac{e^{2}}{\hbar g_{c} B}T^{b}\exp{\left(-\frac{B}
{T^{b}\xi}\right)}
\label{eq:Abrahams_insulator_solu_with_diffu_leng}
\end{equation}
The critical difference between Eq. (\ref{eq:c4}) and Eq. (\ref{eq:Abrahams_insulator_solu_with_diffu_leng}) is linked to the comparison between $L_{in}$ and the hopping length $R_{h}$. When $L_{in}\sim R_{h}$, the conduction is the Mott variable-range hopping determined by $R_{h}$. This is the case of (iv).  As $L_{in}$ becomes less than $R_{h}$, the conduction is non-optimal hopping determined by $L_{in}$. \cite{r11} This is the case of Eq. (\ref{eq:Abrahams_insulator_solu_with_diffu_leng}). Combining (\ref{eq:Abrahams_metal_solu_cond_AA}) and (\ref{eq:Abrahams_insulator_solu_with_diffu_leng}) a unified fitting equation is,  
\begin{equation}
G=G_{0}+G_{A}T^{1/3}+G_{L}T^{b}\exp{\left(-\frac{B_{L}}{T^{b}}\right)}
\label{eq:lowt}
\end{equation}
where $G_{0}$, $G_{A}$, $G_{L}$ and $B_{L}$ are parameters.
 
\begin{figure}[hbt]
	\centering
    \includegraphics[scale=0.4]{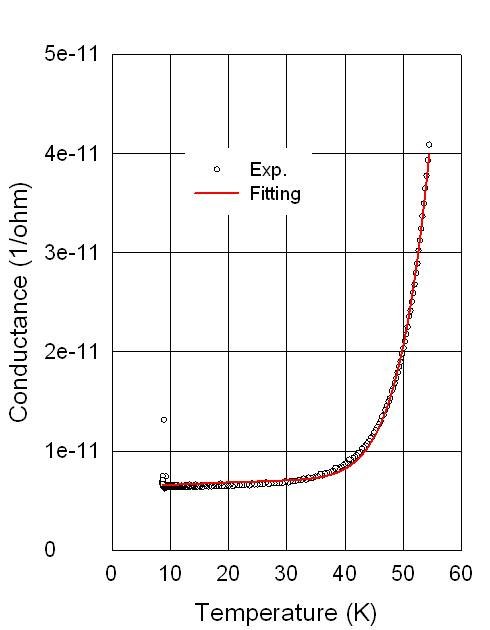} 
	\caption{Conductance vs temperature with the scaling theory,InGaAs: ErAs sample. }\label{fig:lowt}
\end{figure}

The generated fitting from Eq.(\ref{eq:lowt}) agrees well with the experimental curve (Fig. \ref{fig:lowt}). The fitting parameters are listed in Table \ref{table:ta2}.
\begin{table}[h]
\caption{The fitting parameters of the scaling theory} 
\centering 
\begin{tabular}{ c c c c c } 
\hline\hline 
  $B_{L}$  & $b $ & $G_{0}$ & $G_{A}$ & $G_{L}$ \\ [0.5ex] 
\hline 
 197.2 & 0.653 & 5.64E-12 & 4.43E-13 & 4.84E-6\\ 
\hline 
\end{tabular}
\label{table:ta2} 
\end{table}

The value of b is in the proximity of the Efros-Shklovskii exponent of 1/2. This suggests that the conduction mechanism on the insulating side could also be a variable-range hopping type with the Coulomb gap within the density of states. \cite{r11}  The minimum metallic conductivity of InGaAs:ErAs is estimated from its conductance $G_{0}$, $\sigma_{min}\approx 1\times 10^{-10}$ ($\Omega$-cm)\textsuperscript{-1}, ($\sigma_{min}=G_{0}/tN_{s}$, with $t$=2$\mu$m, $N_{s}$=293 \cite{r4}). According to Eq. (\ref{eq:Abrahams_metal_solu_cond_AA}), $\sigma_{min}=e^{2}/\hbar l_{0}$, and the length $l_{0}$ is therefore estimated $\sim 5.3 \times 10^{3}$ cm.  
 
This length- which is unusually large - can be explained with the potential well model. For a three dimensional well with depth $U$ and radius $R$, the localization length or the decay length of ground energy level can be estimated from \cite{ir1}
\begin{equation} 
\xi=\frac{R U_{0}}{(U-U_{0})} \label{eq:locallegn}
\end{equation}  
when $U>U_{0}$, and $U_{0}$ is determined by the radius together with the effective mass of carrier $m$,
\begin{equation}
U_{0}=\pi^{2}\hbar^{2}/8mR^{2}\label{eq:U00}
\end{equation}
\begin{figure}[tbh]
	\centering
    \includegraphics[scale=0.34]{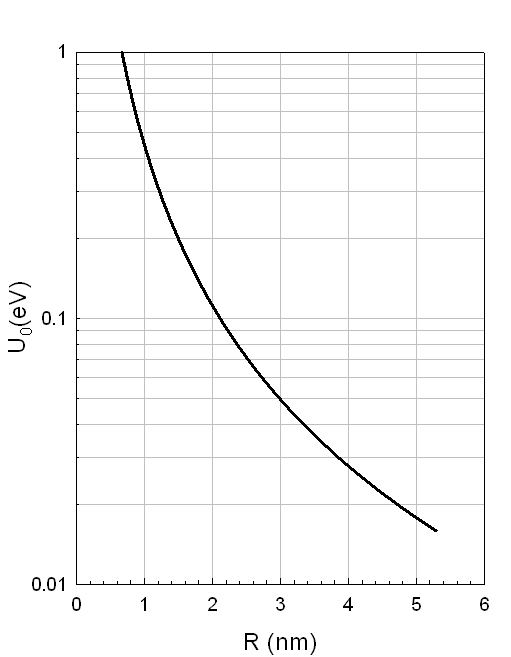}  
	\caption{The estimation of $U_{0}$ vs $R$. Here the effective electron mass is taken as $m=0.21m_{0}$.}	
    \label{fig:quantumwellmodel}
\end{figure}  
According to Eq. (\ref{eq:locallegn}), the localization length $\xi$ becomes significant as the Fermi level and the ground energy $U_{0}$ is resonant with each other. This resonance, $U_{0}=E_{F}$=0.18eV from Table \ref{table:ta1}, occurs at $2R\sim $ 3 nm (Fig. \ref{fig:quantumwellmodel}). Indeed this size is within the range of ErAs particles: 30nm x 4nm to as small as 2nm x 2nm obtained from TEM analysis. As the localization length is sufficiently greater than the scaling length - the  Altshuler-Aronov interaction length - $L_{int}$, the system is metallic. Thus the size of ErAs particles is a critical parameter to the conduction mechanism.

The metallic conductivity in semiconductor nanocomposite isn't an isolated case. We also observed similar saturation behavior in another ultrafast material InGaAsP: Fe. The material was created by high-energy ion implantation into an InGaAsP epitaxial layer lattice-matched to InP and then a optimized anneal \cite{r12}. Subpicosecond photocarrier lifetime has been demonstrated. \cite{r13, r14} However, it remains unclear whether the metallic conduction is from the sub-micron scale InGaAsP microcrystallines in the host InGaAsP layer or the Fe precipitates contained in InP substrate. \cite{r14}

In summary, cryogenic electrical conductance measurements reveal universal conduction processes of a ultrafast semiconductor containing nanoscale ErAs particles from thermal activation near room temperature; to nearest-neighbor hopping variable-range hopping and Anderson localization at intermediate temperature; to metallic behavior at the lowest temperatures. The metallic behavior is explained with the Abrahams scaling theory together with the size effect of quantum potential wells.  
\\
\\
Acknowledgments- This material is based upon work supported by, or in part by, the U. S. Army Research Laboratory and the U. S. Army Research Office under contract/grant number W911NF1210496.  The authors also acknowledge Dr. H. Lu for information on the InGaAs:ErAs sample, and Dr. David Tomich of the Air Force Research Laboratory (AFRL) for providing the undoped InGaAs-on-InP epitaxial sample.


\begin{thebibliography}{1}
\bibitem{r1}M. Sukhotin, E. R. Brown, D. Driscoll, M. Hanson, and A. C. Gossard, Appl. Phys. Lett. 83,  3921 (2003).
\bibitem{r2}D. C. Driscoll, M. P. Hanson, A. C. Gossard, and E. R. Brown, Appl. Phys. Lett.  86, 51908 (2005).
\bibitem{r3}M. Sukhotin, E. R. Brown, A. C. Gossard, D. Driscoll, M. Hanson, P. Maker, and R. Muller, Appl. Phys. Lett. 82, 3116 (2003).
\bibitem{r4}E. R. Brown, K. K. Williams,W.-D. Zhang, J. Suen, Hong Lu, J. Zide and A. C. Gossard, IEEE Trans. Nanotech., 8, 402 (2009).
\bibitem{r6}Elisabeth Muller, Paul Scherrer Institut Wuerenlingen und Villigen, Switzerland.
\bibitem{r7}N. F. Mott and E. A. Davis, Electronic Processes in Non-Crystalline Materials, chap. 6, Clarendon Press, Oxford (1979).
\bibitem{r8}E. Abrahams, P. W. Anderson, D. C. Licciardello, and T. V. Ramakrishman, Phys. Rev. Lett., 10, 673 (1979).
\bibitem{r9}D. J. Newson and M. Pepper, J. Phys. C.: Solid State Phys. 19, 3983(1986). 
\bibitem{r10}B. L. Alshuler and A. G. Aronov, JETP Letter 37, 349, (1983).  
\bibitem{r11}I. P. Zvyagin, Phys, Stat. Sol. (b) 120, 503 (1983).
\bibitem{ir1}C. M. Soukoulis and E. N. Economou, Wave Random Media 9, 255(1999).  
\bibitem{r12}A. Fekecs, M. Bernier, D. Morris, M.  Chicoine, F. Schiettekatte, P. Charette, and R. Ares, Optical Materials Express 1, 1165 (2011). 
\bibitem{r13}M. Martin, E. R. Brown J. Mangeney, A. Fekecs, R. Arés, and D. Morris, Proc. IRMMW-THz 2012.
\bibitem{r14} Private Communication with A. Fekecs, D. Morris and R. Ar\'es from
Institut Interdisciplinaire d’Innovation Technologique (3iT), Universit\'e de Sherbrooke, Sherbrooke, Canada JIK 2R1. 

\end{thebibliography}
\end{document}